# Theoretical study on the possibility of bipolar doping of ScN

G. Soto, M.G. Moreno-Armenta and A. Reyes-Serrato

*Centro de Ciencias de la Materia Condensada, Universidad Nacional Autónoma de México, Apartado Postal 2681, Ensenada Baja California, 22800 México*

**Abstract**

Scandium nitride (ScN) is a semiconducting transition metal nitride for which there are not identified dopants. We present local density functional calculations, in supercell approach, for ScN doped with O and C in N-sites and Ca and Ti in Sc-sites. Small additions of these atoms have the effect of shifting the Fermi level within the electronic band structure. O and Ti bring occupied states in bottom of conduction band, while C and Ca produces holes in top of valence band. Based on the theory we propose that bipolar doping is possible to scandium nitride.

**Introduction**

Ab initio modeling is a very valuable way in which dopant activity can be predicted [1,2]. It is significant that the dopant donor in GaN was first suggested by means of ab initio aids [3]. In the semiconducting material is good to have a good bipolar electrical conduction, i.e. efficient doping from *n*- and *p*-sides. However, due to the reasons which are not yet fully understood, in many materials it is challenging to achieve both, the donors and acceptors dopants. For example, boron is a shallow acceptor in diamond but there are problems relating to shallow donors. Oxygen and silicon are known to be shallow donors in GaN, but still there are difficulties in using efficient shallow *p*-type dopants, Mg and Be are favored but have low activities. Several doping-limiting mechanisms can be cited, of particular importance are: insufficient solubility of the dopant, self-compensation by spontaneous formation of native defects, lattice relaxation around the doping atoms and amphoteric behavior of several potential dopants [1].

In recent years the IIIA-N semiconductors have attracted a great deal of interest because of their wide use in electronics and optoelectronics. The structural and electronic properties of the zincblende and wurtzite phases of BN, AlN, GaN and InN have been studied extensively by different experimental and theoretical groups. Analogous to IIIA-N, the compounds formed by IIIB group metals and N also behave as semiconductors. In this group the most researched is ScN. For long time it was assumed to be a semimetal [4]; however progresses in growing stoichiometrically ScN have undeniably identified it to be an indirect semiconductor with an indirect band gap of about 1 eV [5, 6]. This material has a natural tendency to be *n*-type. In speculatively ways this propensity has been attributed to accidental doping with O, and, or deviations from the 1:1 stoichiometry. Metal vacancies produce *p*-states , as we show in a recent work [7]; however these defects occurs at high energetic costs and probably the crystalline structure would find a compensation mechanism for them by means of lattice distortions. At present it is possible to grown nearly intrinsic ScN, with carrier concentrations in the $10^{17}$ $cm^{-3}$ limits of *n*-type conduction [8]. The electronic structure of ScN is typified by a conduction band with essentially Sc-3*d* character, while the valence band is mostly formed by N 2*p* states. Sc in ScN acts donating its two 4*s* and single 3*d* electrons to N; chemically is similar to an ordinary group IIIA element. The main difference is the existence of disperse *d* states which leads to an indirect band gap with the conduction band minimum at the X point of the Brillouin zone. Until now it is ignored if the presence of these *d* states will have an adverse effect in shallow donors or acceptors.  The aim of this work is to elucidate this point. With that idea in mind, we did a series of calculations, considering C and Ca as electron deficient elements in ScN, and O and Ti as electron donor. There are another possible doping elements not covered here, like Mg and Be.

## Computer Simulation

In order to calculate the electronic structure of undoped and doped ScN we have employed the full potential linearized augmented plane wave (FP-LAPW) method within the framework of density-functional theory, as implemented in the Wien2k code [9]. For structural refinement the exchange-correlation potential was calculated using the generalized gradient approximations in the form given by Perdew-Burke-Ernzerhof (GGA-PBE) [10]. This approximation underestimate the band gap, as a result, ScN comes out to be a nearly zero gap semiconductor in GGA-PBE. Thus, to model the effect of doping is mandatory to add a gap correction. For the density of states (DOS) calculations we applied both the standard GGA-PBE and GGA-Engel-Vosko approximation (GGA-EV). GGA-EV is used as a gap-corrector factor after structural refinements [11]. For doping calculations we have employed the supercell approach with 64 atoms per cell. In our supercells is possible to set 32 metal atoms and 32 nonmetal atoms. The calculated cells were $Sc_{32}N_{32}$, $Sc_{31}Ti_1N_{32}$, $Sc_{31}Ca_1N_{32}$, $Sc_{32}N_{31}C_1$, $Sc_{32}N_{31}O_1$, $Sc_{32}N_{30}C_1O_1$, where the C or O atoms substitute N atoms and Ti and Ca substitute Sc atoms.

## Results and Discussions

The basic shapes of DOS by GGA- PBE and EV calculations agree convincingly within them, although EV opens the band gap of ScN to ~0.5 eV, Figure 1, center plot- still undervalued compared to experimental data [8]. To avoid filtering the effect of the doping agents we used low Gaussian broadenings in plot; this explains the 'noise'. The inserts in graphs show the DOS in the vicinity of band gaps for $Sc_{32}N_{32}$, center plot; $Sc_{32}N_{31}C_1$, upper plot; and $Sc_{32}N_{31}O_1$, lower plot. $Sc_{31}Ca_1N_{32}$ is similar to $Sc_{32}N_{31}C_1$, and $Sc_{31}Ti_1N_{32}$ is similar to $Sc_{32}N_{31}O_1$. As expected, the DOS at the Fermi level ($E_F$) is zero for $Sc_{32}N_{32}$. For $Sc_{32}N_{31}O_1$ and $Sc_{31}Ti_1N_{32}$ $E_F$ is shifted to higher energies (positive value) toward the conduction band. Simultaneously the band gap decreases a shadow amount. The shifting occurs to lower energies for $Sc_{32}N_{31}C_1$ and $Sc_{31}Ca_1N_{32}$. The information of band edges and cohesive energies is resumed in Table 1 for the different cells. The effect of codoping is to cancel the effect of each other, while contracting notoriously the prohibited band, as it is show in table for the $Sc_{32}N_{30}C_1O_1$ entry.

The results show that ScN is behaving like an equivalent IIIA-N semiconductor when the dopants are electron donor or acceptors elements. Keep in mind that these are very high concentrations for dopants (~ $10^{21}$ atoms $cm^{-3}$) and correspond to semiconductors in degenerated states. Our computational resources do not permit lower concentrations. However the important point is clear: the development of Fermi level within the band gap of ScN. We propose that the trends showed here must remain valid even for diluted doping concentrations. Undeniable this is the typical behavior of a doped semiconductor. Another important point is the compatibility of the doping atoms within the semiconductor lattice. Ti and O are in favorable situations, visualized by the differences in cohesive energies in table 1. Thereafter ScN has a natural tendency to be *n*-type. The reaction of ScN with Ti and O are exothermic; the inverse occurs with C and Ca.

**Conclusions**

In summary, the present computational study predicts that O and Ti doping in ScN would lead to *n*-type semiconductor, while C and Ca doping would lead to *p*-type. These results are important given that ScN could be a bipolar semiconductor. There is also a new interesting result of ScN doped with Mn to form a dilute magnetic semiconductor for spintronic applications [12]. A material that can be used simultaneously in electronic and spintronic devices is very worth of a lot of attention for future applications.

**Acknowledgements.** The authors are grateful to C. González, E. Aparicio, J. Peralta and M. Sainz for technical assistance. This work was supported by Departamento de Supercomputo DGSCA-UNAM and Proyecto DGAPA IN120306.

Table 1.- Calculated cohesive energies and band edge parameters.

| | $Sc_{32}N_{32}$ | $Sc_{31}Ti_1N_{32}$ | $Sc_{32}N_{31}O_1$ | $Sc_{31}Ca_1N_{32}$ | $Sc_{32}N_{31}C_1$ | $Sc_{32}N_{30}C_1O_1$ |
|---|---|---|---|---|---|---|
| **$E_{coh}$ (eV)** | -20.97952 | -21.02386 | -21.06251 | -20.55173 | -20.85739 | -20.97976 |
| **$\Delta E_{coh}$ (eV)** | 0 | -0.60302 | -1.12866 | 5.81794 | 1.66097 | -0.00326 |
| **Dos at $E_F$ (states $eV^{-1}$ $Cell^{-1}$)** | 0 | 3.68651032 | 3.8835 | 5.76655006 | 6.37998438 | 0.54639524 |
| **Band Gap (eV)** | 0.5611 | 0.43535 | 0.5578 | 0.54423 | 0.4000 | Absent |
| **Valence band maximum (eV)** | 0 | -0.80546 | -0.9661 | 0.32804 | 0.30028 | - |
| **Conduction band minimum (eV)** | 0.5611 | -0.37008 | -0.4083 | 0.87227 | 0.70029 | - |

**Figures**

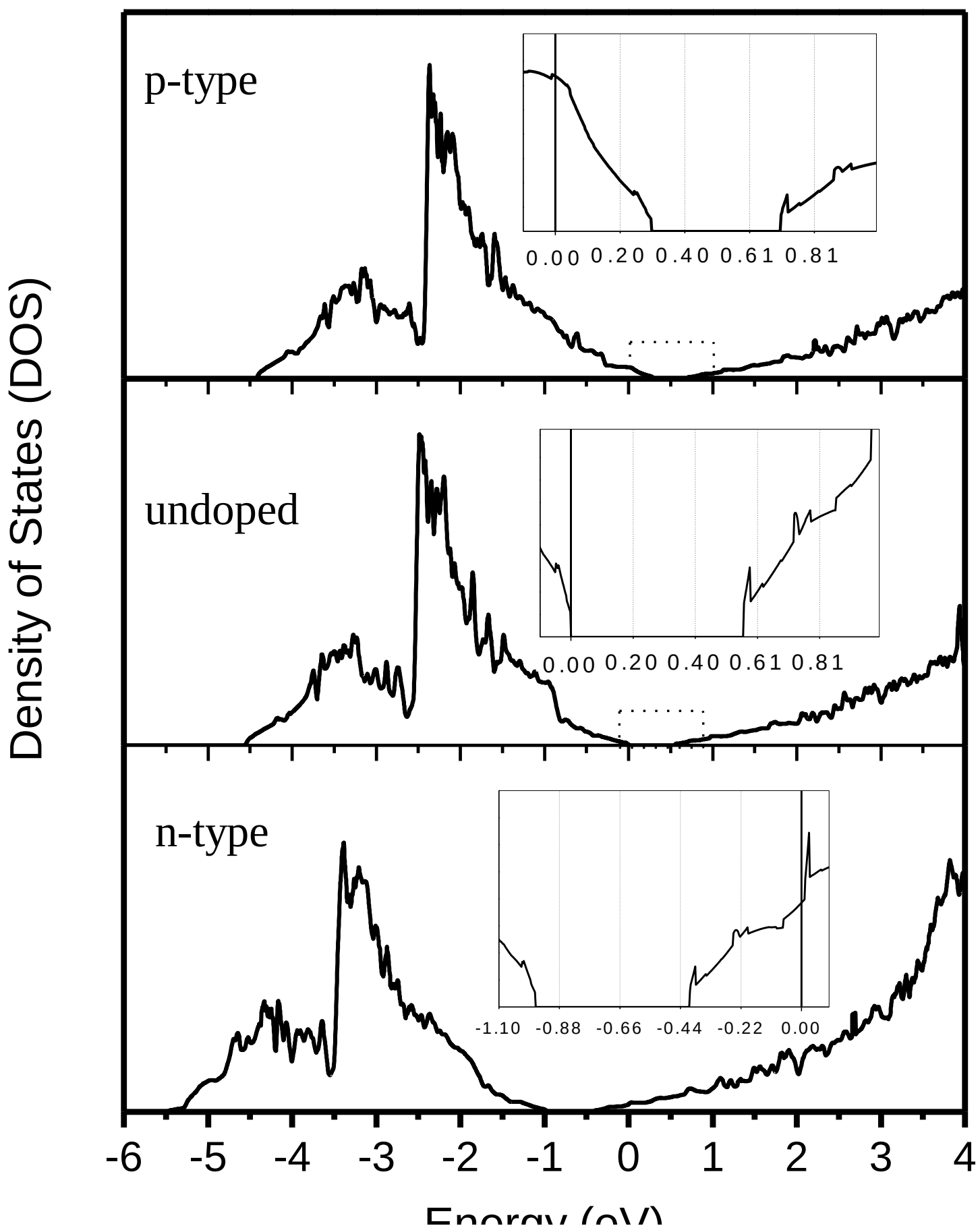

Figure 1.- Density of states of Carbon doped (top), intrinsic (center) and Oxygen doped (bottom) ScN. Inserts are zoomed views in the $E_F$ vicinity.